\documentclass[twocolumn,floats,floatfix,showpacs,amsmath,amssymb,prd,superscriptaddress,nofootinbib]{revtex4-1}

\usepackage{color}

\usepackage{hyperref}
\usepackage{graphicx}
\usepackage{graphics}
\usepackage{epsfig}
\usepackage{bm}
\usepackage{longtable}                 
\usepackage{verbatim}
\usepackage{yfonts}
\usepackage{psfrag}
\usepackage{url}
\usepackage{subfigure}
\usepackage{color}
\input{epsf} 
\usepackage{thumbpdf}
\usepackage{pdfsync}
\usepackage{graphicx}
\usepackage{multirow} 
\def\be{\begin{equation}}
\def\en{\end{equation}}
\def\bea{\begin{eqnarray}}
\def\ena{\end{eqnarray}}

\def\msun{\,{\rm M_\odot}}

\begin{document}

\title{Resolving multiple supermassive black hole binaries with pulsar timing arrays II: genetic algorithm implementation}

\author{Antoine Petiteau} 
\email{petiteau@apc.univ-paris7.fr}
\affiliation{Universit\'e Paris-Diderot-Paris7 APC - UFR de Physique, B\^atiment Condorcet ,10 rue Alice Domont et L\'eonie Duquet 75205 PARIS CEDEX 13, France}
\affiliation{Albert Einstein Institute, Am Muhlenberg 1 D-14476 Golm, Germany}
\author{Stanislav Babak} 
\email{Stanislav.Babak@aei.mpg.de}
\affiliation{Albert Einstein Institute, Am Muhlenberg 1 D-14476 Golm, Germany}
\author{Alberto Sesana} 
\email{Alberto.Sesana@aei.mpg.de}
\affiliation{Albert Einstein Institute, Am Muhlenberg 1 D-14476 Golm, Germany}
\author{Mariana de Ara\'ujo} 
\email{marianabdaraujo@gmail.com}
\affiliation{Albert Einstein Institute, Am Muhlenberg 1 D-14476 Golm, Germany}

\date{\today}

\begin{abstract}
Pulsar timing arrays (PTAs) might detect gravitational waves (GWs) from massive black hole (MBH) binaries 
within this decade. The signal is expected to be an incoherent superposition of several nearly-monochromatic 
waves of different strength. The brightest sources might be individually resolved, and the overall deconvolved, at least partially, in its individual components. 
In this paper we extend the maximum-likelihood based method developed in \cite{babaksesana12}, to search
for individual MBH binaries in PTA data. We model the signal as a collection of circular monochromatic binaries,
each characterized by three free parameters: two angles defining the sky location, and the frequency. 
We marginalize over all other source parameters and we apply an efficient multi-search genetic algorithm to 
maximize the likelihood function and look for sources in synthetic datasets. On datasets characterized by white
Gaussian noise plus few injected sources with signal-to-noise ratio (SNR) in the range 10-60, 
our search algorithm performs well, 
recovering all the injections with no false positives. Individual source SNRs are estimated within few \% of 
the injected values, sky locations are recovered within few degrees, and frequencies are determined with 
sub-Fourier bin precision. 
\end{abstract}


 

\maketitle

\section{Introduction}
\label{S:intro}

Precision timing of millisecond pulsars provides a unique opportunity to get the very first low-frequency gravitational wave (GW) detection. 
This prospect is attracting the attention of the wider astrophysical community, causing a recent boost of activity in the field. 
The Parkes Pulsar Timing Array (PPTA, \cite{manchester2008}), the European Pulsar Timing Array (EPTA, \cite{janssen2008}) and the North American 
Nanohertz Observatory for Gravitational Waves (NANOGrav, \cite{Jenet:2009hk}), joining together in the International Pulsar Timing Array (IPTA, \cite{hobbs2010}), 
are collecting data and improving their sensitivity in the frequency range of $\sim10^{-9}-10^{-6}$ Hz. In the coming years, the Chinese five hundred meter 
aperture spherical telescope (FAST, \cite{smits09}) and the planned Square Kilometer Array (SKA, \cite{lazio2009}) will provide a major leap in sensitivity. 
Current surveys are already placing interesting upper limit on the level of a putative GW background \cite{vanh11,demorest12}, 
skimming the range predicted by state of the art models of MBH evolution \cite{sesana08,sesana09} . Within the next few years, the combined IPTA data might either result in a first detection, or start placing interesting limits on the MBH binary formation efficiency in massive galaxies.

The detection principle is very simple: GWs affect the propagation of radio signals from the pulsar to the receiver on Earth, leaving a characteristic 
fingerprint in the time of arrival of the radio pulses (e.g., \cite{sazhin78,hellings83}). Such fingerprint depends on the properties of the underlying 
cosmological population of inspiralling binaries, and will consists of a superposition of quasi-monochromatic waves, similar to the white dwarf-white dwarf 
foreground (e.g., \cite{Nelemans01}) in the mHz window relevant to space based interferometry \cite{amaro2012}. This signal has generally been regarded as a stochastic 
background, and data analysis techniques has been developed accordingly \cite{jen05,jen06,anholm09, vanh09, vanh11,demorest12}. The actual expected signal, however, 
is far from being isotropically distributed in the sky, with just few sources dominating the power at each frequency \cite{sesana08,sesana09,KS11}. The possibility 
of resolving an individual source offer appealing astrophysical prospects, and PTA capabilities 
on this front were also investigated in details by many authors \cite{jen04,SesanaVecchio10,corbin2010,yardley10,lee11,ellis12}.

What is still missing, is a detailed study of what kind of information a PTA can extract out of a complex superposition of multiple sources. Is the signal going 
to be similar to a confusion noise? Can we resolve individual sources? How many of them? Can we locate them in the sky, and to what level of accuracy? All these 
questions are of great interest for the astrophysical community; precise sky localization of individual sources will allow the efficient search for electromagnetic 
counterparts \cite{sesana12,tanaka12}, opening the new horizon of multimessenger astronomy.

This is a second in a series of paper devoted to the exploration of the PTA potential of resolving multiple GW sources. In \cite{babaksesana12} (hereinafter Paper~I), 
we demonstrated PTA efficiency in disentangling monochromatic sources at the same frequency. The key idea is to estimate the likelihood that a certain number of 
sources with certain parameters are present in the data. We developed a formalism that allowed us to maximize analytically the likelihood over the extrinsic source 
parameters, restricting the search to the source sky location only (2$\times N$ parameters, where $N$ is the number of GW sources in the template). There, we did 
not implement any proper algorithm to search over the parameters space, and we made a lot of simplifying assumptions, suitable to a first, exploratory investigation.

Our aim is to implement a proper search algorithm, progressively relaxing our limiting assumptions to develop a detection pipeline able to handle the whole 
complexity of a realistic dataset. We start in this paper with two major steps: (i) we extend our mathematical formalism to include frequency scan and (ii) we 
present an upgraded version of the genetic algorithm employed by \cite{Petiteau:2010zu}  in the LISA mock data Challenge \cite{Babak:2008aa, Babak:2009cj}   
specifically developed to search for a global maximum on the multimodal likelihood surface embedded in the multidimensional parameter space.
We have found (similarly to the mock LISA data challenge) that the genetic algorithm (GA) is very efficient in 
finding the correct number of sources and their parameters. 

The paper is organized as follows. In Section 2, we spell out our main assumptions and in Section 3 we present the genetic algorithm and its feature. The datasets 
used to test the algorithm are detailed in Section 4, and the algorithm performances and results are presented in Section 5. In Section 6 we draw our conclusion  
and discuss improvements we will present in future work.  

\section{Detection strategy, extension to frequency search}
\label{S:analytics}

The main purpose of this paper is to extend our formalism to include search in frequency and to implement a proper search algorithm to identify maxima in 
the likelihood. Accordingly, we relax number 1, 3 and 8 from the limitations and assumptions 
described in Section II of Paper~I, i.e.:
\begin{enumerate} 
\item we consider only datasets with noise;
\item we inject sources at different frequencies; 
\item we implement a proper search algorithm to maximize the likelihood. 
\end{enumerate}

\subsection{Choice of the template}

In computing the likelihood function, we consider monochromatic GW sources assuming that the orbital frequency does not 
change appreciably over the observation period (see, e.g., \cite{SesanaVecchio10}) which we took to be 10 years. Each GW signal is therefore characterized by seven parameters only: the overall signal amplitude $\{{\cal A}\}$, the source frequency and phase $\{f, \Phi_0\}$, and the angles defining its location in the sky $\{\phi, \theta\}$, inclination $\{\iota\}$, and polarization $\{\psi\}$. Contrary to Paper~I, we do not fix all the systems at the same frequency but we consider the unknown frequencies of the sources as additional search parameters. 

We do not use the full response of
the pulsar-Earth detector, which is combined out of the ``Earth term'' and the ``Pulsar term'' (see, e.g., \cite{SesanaVecchio10}) (from now on we drop the quotes and use those notions as a jargon), but we construct a signal template to match the Earth term only. There are several reasons to 'drop' the pulsar term in the analysis.
In all pulsars, the Earth terms add-up coherently: they all have the same frequency and phase, and the amplitude of the signal in the residuals depends on the relative position of the pulsar and the GW source on the sky. Conversely, the pulsar terms are in general incoherent:
they usually appear at different frequencies, and the phase and amplitude of the signal depends not only on the position of the source relative to the pulsar,
but also on the distance to the pulsar which is usually poorly known (in most of the cases to $\sim 10$\% precision). Even if we assume we know the pulsar distance exactly,
the pulsar term carries the imprint of the binary system emitting as it was a time $\Delta t = L(1 + \hat{k}.\hat{n})$ in the past as compared to the Earth term; where $L$ is the Earth-pulsar distance and $\hat{k}$ and $\hat{n}$ are the unit vectors pointing to the sky location of the source and the pulsar respectively. This means that to connect
pulsar and Earth terms we need to know the evolution of the binary system for $\Delta t$ which is typically $10^3-10^4$ years. Even assuming pure GW evolution, the prediction of the signal at the pulsar term will be affected by spin-orbit coupling precession \citep{mingarelli12}, or non negligible eccentricity (which is very likely for broad binaries (see, e.g., \cite{sesana10,preto11}). In addition to this, the 
assumption that the binary evolution is driven by GW back reaction only could not hold for such system. Widely separated MBH binaries could still dynamically interact with the surrounding gas and/or stellar environment, which might significantly affect (and sometimes dominate) their orbital evolution \cite{KS11}. In other words, the inclusion of the pulsar term is {\it always} model dependent, and we try to avoid it. Considering the Earth term only also has some drawbacks. If the
systems evolve only under GW radiation reaction, then relatively light and wide binaries might not have evolved much (less than a Fourier bin)
over the pulsar-Earth light travel time; in this case the pulsar and Earth terms appear effectively at the same frequency and we need to take that into account. Moreover, pulsar terms from different GW sources will inevitably overlap in frequency with the Earth term of a given source, creating a spurious contribution to the signal. Here we inject in the data only the Earth portion of the signal, and we delegate problems related to pulsar terms to future work.


\subsection{Likelihood function and detection statistics}

The details of the detection statistics were outlined in Paper~I, we 
briefly summarize here the main points and describe the extension of the formalism to sources with different frequencies. As justified in the previous section, we use the matched filtering technique assuming the Earth term as a template. The mathematical description of the signal template for an individual source as a function of the parameters $\vec{\lambda}=\{{\cal A}, f, \Phi_0, \phi, \theta, \iota, \psi\}$ is given by equations (11)-(16) of Paper~I. The log-likelihood ratio (likelihood that a dataset $x_\alpha(t)$ contains a GW signal $r_{\alpha}(t; \vec{\lambda})$ over the likelihood that it is pure noise), is 
\begin{equation}
\log{\Lambda} = <x_{\alpha} | r_{\alpha}> - \frac1{2}<r_{\alpha} | r_{\alpha}>,
\label{loglambda}
\end{equation}
where the subscript $\alpha$ corresponds to a given pulsar and $r_{\alpha} = r^E_{\alpha}$ is the expected Earth term in the data. We neglect here
all possible stochastic GW signals and look  for individual binaries standing above a putative unresolved background only. The inner product appearing in equation (\ref{loglambda}) is defined as
\begin{equation}
<x | r> =  \frac{2 T_o}{N S(f)}\sum_{i = 1}^{N} x(t_i) r(t_i),
\label{inner}
\end{equation}
where $N$ is the number of points in the time series, $T_o$ is the observation time, and, $S(f)$ is one-sided noise power spectral density which we assume to be white Gaussian. Equation (\ref{inner})
is the discrete version of the inner product used in Paper~I; it has the advantage to be applicable to unevenly sampled data, which will be the case in reality.
It was shown in Paper~I, that the GW signal imprinted in the data by each individual source can be written as
\begin{equation}
r_{\alpha} = \sum_{j=1}^4  a_{(j)} h_{(j)}^{\alpha},
\end{equation}
where $h_{(j)}$ are time dependent functions that include the parameters we want to search for, while $a_{(j)}$ are constants over the observation period. 
Expressions for $h_{(j)}$ and $a_{(j)}$ are the same as given in Paper~I. Here, we search over sky location and frequency $f$ of each GW signal, but we assume 
$f$ to be constant over the time of observation, therefore we can keep it either as part of $a_{(j)}$  (as in the Paper~I) or we can include it in $h_{(j)}$.
In practice we decompose the signal in  $h_{(j)}(\theta,\phi,f)$ and $a_{(j)}({\cal A}, \Phi_0, \iota, \psi)$ (therefore shifting the $f$ dependence from $a_{(j)}$ to $h_{(j)}$ compared to the expressions given in Paper~I).  

We can then maximize the likelihood ratio over the $a_{(j)}$ constants for each GW source analytically:
\bea
\frac{\partial \log(\Lambda)}{\partial a_{(j)}} = 0,\;\;\;\; \to a_{(k)} = M^{-1}_{kj}X_j,
\label{E:MLE}\\
\{ \log(\Lambda) \}_{{\rm max}\{a_{(j)}\}} \equiv {\mathcal F}_e = \frac1{2} X_k M^{-1}_{jk} X_j,
\label{E:fstat}
\ena
where
\bea
\label{eq:XM}
X_{j} \equiv \sum_{\alpha=1}^P <x_{\alpha} | h^{\alpha}_{(j)}>,\;\;\;\;   M_{jk} \equiv \sum_{\alpha=1}^P
<h_{(j)}^{\alpha} | h_{(k)}^{\alpha}>.
\ena
The statistical properties of ${\mathcal F}_e$ are investigated in details in \citep{ellis12}.
In presence of $N_s$ GW sources in the template, the coefficients $a_{(j)}$ are represented 
by a $4\times N_s$ array,
$X_j$ is also a $4 \times N_s$ array, while the $M$-matrix is  a $4N_s \times 4N_s$, 2-D matrix.
The matrix can be decomposed in $N_s$ $4\times 4$ row- and column-matrices, each corresponding
to the cross terms between the $I$-th and $J$-th GW sources:

\bea
M^{IJ} = \sum_{\alpha} \left( \begin{array}{cccc}
U^{IJ} \mathcal{I}_{ss} ^{IJ} &  Q^{IJ} \mathcal{I}_{ss}^{IJ} & U^{IJ} \mathcal{I}_{sc} ^{IJ} &  Q^{IJ} \mathcal{I}_{sc}^{IJ} \\
Q^{IJ} \mathcal{I}_{ss} ^{IJ} &  V^{IJ} \mathcal{I}_{ss}^{IJ} & Q^{IJ} \mathcal{I}_{sc} ^{IJ} &  V^{IJ} \mathcal{I}_{sc}^{IJ} \\
U^{IJ} \mathcal{I}_{cs} ^{IJ} &  Q^{IJ} \mathcal{I}_{cs}^{IJ} & U^{IJ} \mathcal{I}_{cc} ^{IJ} &  Q^{IJ} \mathcal{I}_{cc}^{IJ} \\
Q^{IJ} \mathcal{I}_{cs} ^{IJ} &  V^{IJ} \mathcal{I}_{cs}^{IJ} & Q^{IJ} \mathcal{I}_{cc} ^{IJ} &  V^{IJ} \mathcal{I}_{cc}^{IJ} 
\end{array}\right)
\ena
where
\bea
U^{IJ} &=& (F^{\alpha}_c)^I  (F^{\alpha}_c)^J, \;\;\;  Q^{IJ} =  (F^{\alpha}_c)^{I} (F_s^{\alpha})^{J}, \;\;\; \nonumber\\
V^{IJ} &=& (F^{\alpha}_s)^I(F^{\alpha}_s)^J,
\ena
and $F^{\alpha}_{c,s}$ represent the decomposition of the antenna pattern given by equation (16) of Paper~I. The $\mathcal{I}$  terms come from the inner products of the time dependent
parts of $h_{(j)}^{\alpha}$ which, for each source $I$ are cos- and sin-functions of phase $\phi_I = 2\pi f_I t \equiv \omega_I t$. We can evaluate those inner products analytically 
by using the integral representation adopted in Paper~I; for example:
\bea
< h^I_{\alpha, (1)} &|& h^J_{\alpha, (1)} > =  (F^{\alpha}_c)^I  (F^{\alpha}_c)^J <\sin(\phi_I) | \sin(\phi_J)  > \nonumber \\
& & \sim (F^{\alpha}_c)^I  (F^{\alpha}_c)^J \frac{2}{T_o} \int_0^{T_o} \sin(\phi_I) \sin( \phi_J)\; dt \nonumber \\
& & \equiv (F^{\alpha}_c)^I  (F^{\alpha}_c)^J\mathcal{I}_{ss} ^{IJ} \equiv U^{IJ} \mathcal{I}_{ss} ^{IJ}.
\ena
The explicit form of the $\mathcal{I}$ integrals for all possible sine and cosine combinations are given by:
\bea
\mathcal{I}_{ss}^{IJ} &=& \frac{2}{T_o} \int_0^{T_o} \sin(\omega^I t) \sin( \omega^J t)\; dt   \nonumber \\
 & = &  {\rm sinc}(\Delta \phi) - {\rm sinc}(\Sigma \phi)  \\
\mathcal{I}_{cc}^{IJ} &=& \frac{2}{T_o} \int_0^{T_o} \cos( \omega^I t) \cos( \omega^J t)\; dt \nonumber \\
 & = & {\rm sinc}(\Delta \phi) + {\rm sinc}(\Sigma \phi) \\
\mathcal{I}_{sc}^{IJ} &=& \frac{2}{T_o} \int_0^{T_o} \sin( \omega^I t) \cos( \omega^J t)\; dt   \\
 & = & \sin\left(\frac{\Sigma \phi}{2}\right) {\rm sinc}\left(\frac{\Sigma \phi}{2}\right) + \sin\left(\frac{\Delta\phi}{2}\right){\rm  sinc}\left(\frac{\Delta\phi}{2}\right) \nonumber \\
\mathcal{I}_{cs}^{IJ} &=& \frac{2}{T_o} \int_0^{T_o} \cos( \omega^I t) \sin( \omega^J t)\; dt   \\
 & = & \sin\left(\frac{\Sigma \phi}{2}\right) {\rm sinc}\left(\frac{\Sigma\phi}{2}\right) - \sin\left(\frac{\Delta\phi}{2}\right)  {\rm sinc}\left(\frac{\Delta\phi}{2}\right), \nonumber
\ena
where $\Delta\phi \equiv (\omega^I - \omega^J)T_o$ and $\Sigma\phi \equiv (\omega^I + \omega^J)T_o$. Note that the $M$ matrices reduce to the expression given in equation (25) of Paper~I when $\omega^I = \omega^J$. The $I\ne J$ terms give beatings between two signals at different frequencies and they are usually smaller than the terms in the $I=J$ matrices. We found that one can consider the sources approximately at the same frequency if $|f_I - f_J|  \approx (2/3) \Delta F$, where $\Delta F = 1/T_o$ is the size of the Fourier frequency bin.

We use ${\mathcal F}_e$ as detection statistic. Note that we can also estimate the relative contribution of each source as  ${\mathcal F}_e^J =  \frac1{2} X_k^J (M^{-1}_{jk})^J X_j^J$. Following \cite{lrr-2012-4}, we can express the relation between the expectation of the  analytically maximized likelihood ${\mathcal F}_e$  and the SNR as follow:  
\be
E({\mathcal F}_e ) = \frac1{2}\left( 4N_s + {\rm SNR}^2\right).
\en
To search for an individual source we use the same mathematical framework assuming $N_s = 1$, we refer to \cite{ellis12} for more details on the statistical properties of ${\mathcal F}_e$ in this latter case.

\section{Multi-search genetic algorithm:  description and implementation}
\label{S:gMBenetic}

 We search for the maximum of ${\mathcal F}_e$ with a modified version of the genetic algorithm (GA) described in~\cite{Petiteau:2010zu}, performing multiple searches in parallel. 

\subsection{Genetic algorithm}

\begin{table*}[htdp]
\begin{center}
\begin{tabular}{ccc}
Genetic algorithm & & GW search \\
\hline
organism & $\Longleftrightarrow$ & template : signal from $N_s$ GW sources\\
gene (of an organism) & $\Longleftrightarrow$ & parameter (of a template) : $3 \times N_s$ \\
allele (of a gene) & $\Longleftrightarrow$ & bits (of the value of the parameter) \\
quality $Q$ & $\Longleftrightarrow$ & Maximized Likelihood, i.e. $F$-statistic ${\mathcal F}_e$\\
colony of organisms & $\Longleftrightarrow$ &  evolving group of templates \\
$n$-th generation & $\Longleftrightarrow$ & the state of colony at $n$-th step of evolution \\
(selection + breeding) + mutation & $\Longleftrightarrow$ & w parameter space exploration strategy
\end{tabular}
\end{center}
\caption{Correspondence between GA and GW data analysis notions.}
\label{T:DicoGWsearchGA}
\end{table*}%

The GA is a method to perform global searches on large parameter spaces 
(optimization method) based on the natural selection principle. 
In nature, organisms adapt themselves to their environment: the
smartest/strongest/healthiest organisms are more likely to survive
and participate in the breeding to produce the offspring. 
These two processes, selection and breeding, are used in GAs
to produce subsequent generations of organisms. Since the best organisms are more
likely to participate in breeding, the new generation should be
better (in which sense we will specify later) than the previous one,
or at least no worse. This procedure leads to an evolution of the 
organism population, just like in nature: the
good qualities of the parents can be transferred to their offspring. 
In the biological world, besides  these two basic operations, among every generation, 
there are always  few individuals which have better characteristics to adapt to the environment,  
produced as a result of positive mutations. By introducing new genotypes into the 
population, mutations can potentially improve the forthcoming generations and 
consequently accelerate the evolution towards a perfect adaptation to the environment.

We apply these principles to the search for individual GWs in PTA data using equivalences 
described in table~\ref{T:DicoGWsearchGA}: a template described by a set of parameters 
$\{ \theta_I , \phi_I, f_I \}$ is one organism described by a set of genes; the 
${\mathcal F}_e$ of the template is the quality of the organism; the binary representation 
of a parameter by a set of bits is the representation of a gene by a set of allele.

We start with a group of organisms (templates) chosen randomly (initial search) 
or constructed from the results of previous searches. 
We evaluate the quality of each organism (${\mathcal F}_e$).
We select set of pairs (parents) based on their quality: organisms with better
quality (templates with higher  ${\mathcal F}_e$) are chosen more often than weak organisms.
We combine the genotypes of two parents to produce a child (we combine parameters
of two chosen templates to produce a new one). We impose the number of produced
children  to be equal to the number of parents (i.e., we keep the number of evolving organisms constant at each generation).
Next, we allow  with a certain probability a random mutation in the children's genes
(with some probability we randomly change the parameters of the new templates, 
exploring a larger area of the parameter space). The parents are discarded and 
the resulting children form a new generation. We repeat 
the procedure until we reach a steady state (maximum in the quality), 
or up to a maximum number of generations. We keep only one generation active 
(one group of templates).

We now turn to describe the tree steps used for making a new generation 
(replacing parents by children), specifying the used possibilities of tuning for each process.   

In the {\it selection process}, two parent  organisms are selected. The probability to chose an organism is directly related to its
 quality  ${\mathcal F}_e$: the higher is ${\mathcal F}_e$ of an organism, the higher is the probability to be selected. The  
relation between ${\mathcal F}_e$ and the selection probability is moderated by a parameter called
 "temperature". The higher is the temperature, the smaller is the difference in selection probability among the organisms (for infinite
 temperature, all probabilities are equal); the lower is the temperature, the higher is the probability to select only the best
 organisms. The value of the temperature evolves during the search (similar to simulated annealing  \cite{Cornish:2006ms}): it starts at high temperature and then decreases and alternates between hot and cold phases: i.e., we allow at the beginning a large exploration range, by accepting good and bad organisms, but later on we search only around the better ones, allowing some jumps (by alternating hot and cold phases).
More details about the selection criteria can be found in sections III.C and IV.A.2 of~\cite{Petiteau:2010zu}.

In the {\it breeding process}, the parameters of the two selected templates are mixed by combining the bits of their binary 
representation, i.e. by combining the allele of the genes. More details on the breeding and code of the genes can be found in sections
III.B and III.D of~\cite{Petiteau:2010zu}. At each generation, we change the binary representation alternating between
 "standard binary" and "gray code". The breeding method used here is the "crossover one random point" 
 which consists of taking the first bits from one parent and the last ones from the other ; the cross point is chosen randomly.    
   
In the {\it mutation process}, some of the bits are changed. The mutation rate is managed by a parameter called PMR 
(Probability Mutation Rate). At the beginning of a search, a gene is mutated with a probability described by the PMR.
Then for mutating a gene, 8 bits of this gene are (over 20) randomly chosen and changed. 
After a certain number of generations (typically 300), the type of mutation is changed : 
each bit of each gene is directly mutated with a probability at PMR. 
The value of the PMR decreases during the search, starting around 0.5 - 0.1 and ending around 0.1 - 0.01.  
Using these two types of mutation and the PMR evolution corresponds again to start with a large exploration, slowly shrinking to the area around the best templates only at late generations.
More details on the mutation are provided in sections III.E and IV.A.3 of~\cite{Petiteau:2010zu}.

In addition, we always reproduce the best organism between two generations (elitism - cloning of the best). 
This means that the algorithm always converges toward the best solution. 
Finally, we also use the local mutation described in section IV.B.2 of~\cite{Petiteau:2010zu}.

We typically use 50 organisms per generation and 1000 generations. 
The run of one GA takes few minutes on a standard laptop (one Intel core at 2GHz).
Since the size of the parameters space increases with the number of sources in the template, 
the convergence speed decreases accordingly.
The algorithm usually converges around the true solution in less than 400 generation for the highest SNR sources.
One of the most interesting features of the GA is its efficiency in finding maxima in the $\mathcal{F}_e$ surface first (during the large exploration phase),
and then in exploring them deeply to extract the global one (during the local exploration phase).
One GA run is usually sufficient to find most of the sources, but sometime it gets stuck on some local maximum. To overcome this problem, we run several GAs in parallel, as described in the next section.

\subsection{Multiple searches (MultiSearch)}

The GA described in the previous section provide the basis for a more general method called "multiple searches"  (MS) algorithm. This method consists in running several GAs in parallel with different properties and initial parameters.

We take an initial population with parameters chosen randomly. We start a GA on this population, 
tuning the parameters to perform a large exploration.
In the resulting population, we select only the best organism which are well separated. 
This means that the selected organisms have 
SNR $> 97 \%\, $SNR$_{\rm Best}$ and the distance in parameter space between two organisms is higher than a 
certain threshold chosen empirically after a number of tests: 
$| \cos(\theta_{I,i}) - \cos(\theta_{I,j}) | > \Delta_{c\theta} = 0.1$ , $| \phi_{I,i} - \phi_{I,j} | > \Delta_{\phi} = 20^o$ and 
 $| f_{I,i} - f_{I,j} | > \Delta_{f} = 0.5  \text{nHz}$,  where $I$ refers to the source and $i$ and $j$ to the solutions.
The selected solutions are called "modes" and this selection process is called "mode separation".

\begin{figure*}
\centering
\begin{tabular}{cc}
\includegraphics[width=0.44\textwidth, keepaspectratio=true,]{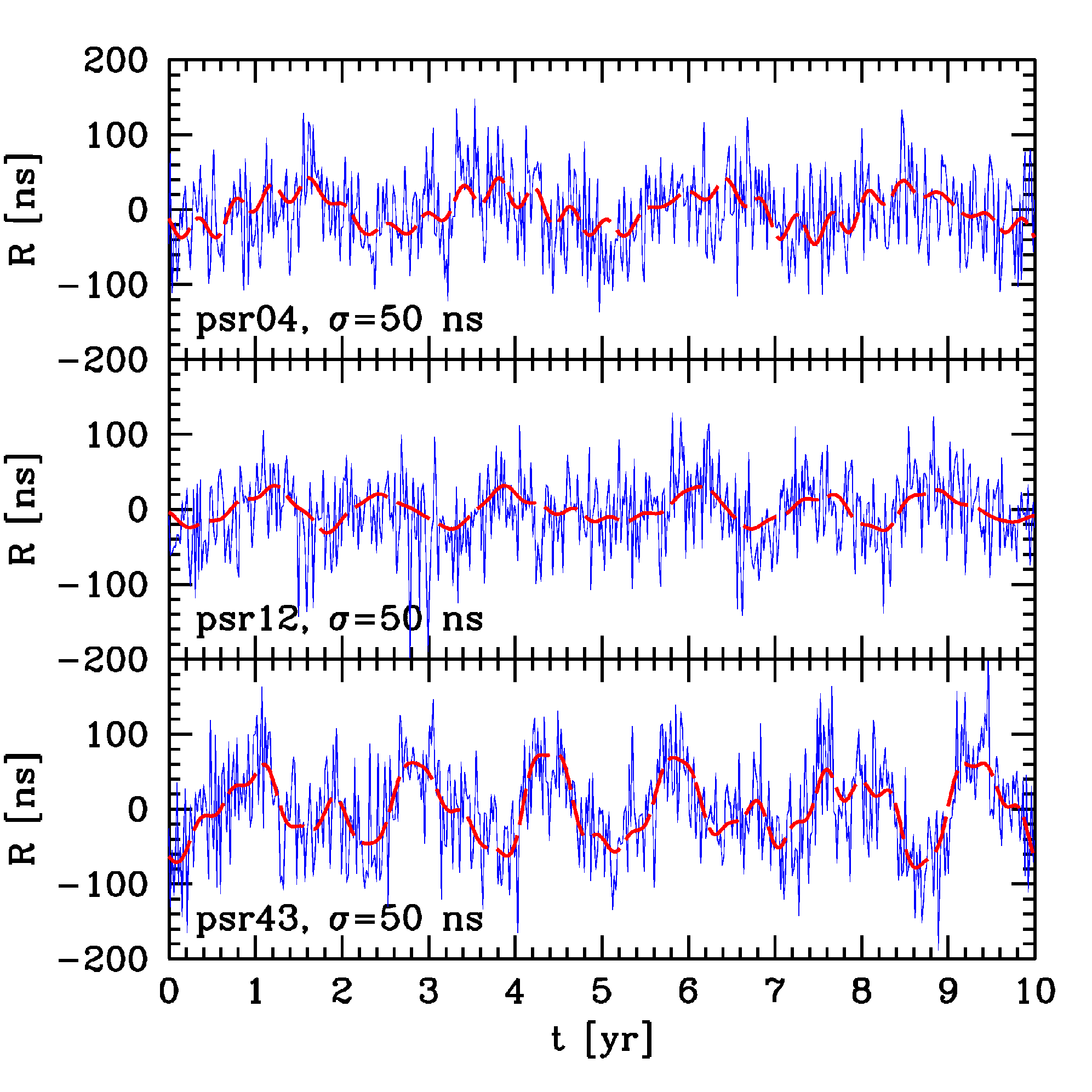} &
\includegraphics[width=0.44\textwidth, keepaspectratio=true,]{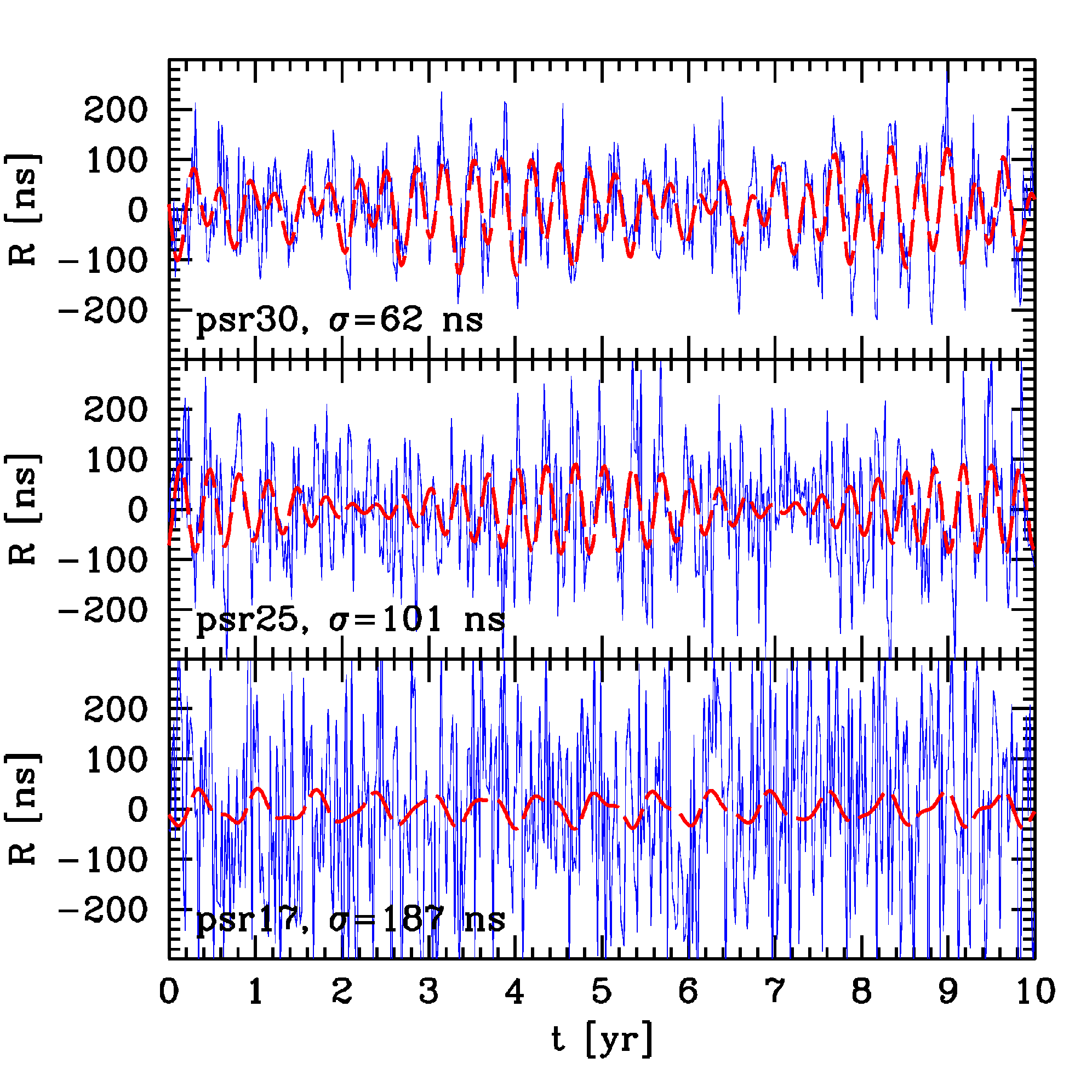} 
\end{tabular}
\caption{Sample of simulated timeseries. In each panel, the red dashed curve is the injected signal, where the blue jagged line represent the total raw 
dataset including signal plus white Gaussian noise. Left panel: pulsars extracted from Dataset3; sources are injected in white Gaussian noise with 
rms $\sigma=50$ns. Right panel: pulsars extracted from Dataset4; here each pulsar has a different noise level, as labelled in each panel.}
\label{timeseries}
\end{figure*} 

The next step is to start one GA on each mode, tuned for local exploration. 
The goal is to explore the vicinity of the mode to find the local highest value of ${\mathcal F}_e$. 
The organisms of each GA, are allowed to explore only their mode neighborhood and are forbidden to go on the area of interest of other modes.
The area of interest of a mode $\{ \cos\theta_{i}, \phi_{i}, f_{i}\}$ is defined within
 $[\cos\theta_{i}- \Delta_{c\theta}, \cos\theta_{i}+ \Delta_{c\theta} ]$,
 $[\phi_{i}- \Delta_{\phi}, \phi_{i}+ \Delta_{\phi} ]$ and
 $[f_{i}-\Delta_{f},f_{i}+\Delta_{f}]$.
In parallel to these 'mode GAs', we start another GA tuned for large exploration. We forbid the 
organisms of this GA to go on the areas of the modes. The aim of this GA is to find new modes (overlooked
in previous searches), if there are any left (it can also give a null result).

At the end of this step, all the solutions are grouped together and we apply the "mode separation" 
to identify "modes". Then we iterate the procedure by restarting several 'local' GAs.

In the long run, this method, as other stochastic methods (e.g., Markov Chain Monte Carlo methods), 
is guaranteed to converge to the global maximum. However, there is no way to exactly know a priori how
fast it will do so, and one has to decide when to stop it, being somehow confident that the 
best solution has been found. 
We usually do 2 to 5 iterations of the procedure outlined above before stopping. 
The number of modes $N_{\rm modes}$ found increases with the number of iterations.
Since we are running $N_{\rm modes}+1$ GAs at each iteration, the first one takes just few minutes (one initial exploratory GA run), the second one can take up to an hour (depending on the number of modes found in the first iteration), and the later ones up to few hours. In total, we run between 50 to 300 GAs for a search. The correct solution is usually found after 2 iterations (i.e. about one hour). As a pseudo-test for convergence, we run several times (typically 10) our MS-GA code, with different initial conditions. If all the run give almost the same results, we claim convergence.

\section{Description of the test datasets}
\label{S:datasets}

The genetic algorithm described in the previous section was used to analyse four blind datasets, 
which we describe here in detail. Each dataset consists of a 
collection of time series representing the residuals obtained by timing an ensemble of millisecond pulsars (MSPs). 
In all datasets, MSPs are placed randomly in the celestial sphere, 
each time series consists of 523 equally sampled datapoints over a total observing time of 10 
years (one datapoint every two weeks), and the noise is assumed to 
be white Gaussian. The injected sources were all equal mass, circular, non spinning binaries 
with chirp mass of $10^9\msun$, placed at the same redshift 
(distance), but with sky location, inclination, polarization and initial phase drawn randomly, resulting in a range of signal strengths. 
The redshift was chosen to produce the desired SNR range in the 
datasets (see below). The frequency was drawn from a random distribution in the range $10^{-8}-10^{-7}$Hz. 
Sources were evolved according to equation of motion accurate to 3.5 Post Newtonian order in phase evolution 
\cite{junker92}, and gravitational waveforms were generated following \cite{SesanaVecchio10} (see also \cite{bc04,wvk08}). The final residual injected in the datasets were obtained by time integration of the waveforms
(see equations (8) and (9) of Paper~I). Note that the injected data are quite different from the adopted
circular, non evolving monochromatic templates we use in the search; we are therefore mimicking the 
(likely) situation in which the template does not perfectly match the signal. 
Especially at high frequency, there might be a non-negligible evolution of the source frequency over 
10 years, possibly introducing a bias in our source recovery. We will quantify this effect in our results. 
As in Paper~I, we considered the Earth term only (issues related to pulsar terms will be explored in 
the next paper). Here follow the details of the four datasets:

\begin{itemize}
\item {\bf Dataset1}: 30 MSP; rms noise $50$ns in each pulsar; 5 binaries at $z=0.01$, with individual SNR in the range $\sim 30-60$;
\item {\bf Dataset2}: 30 MSP; rms noise $50$ns in each pulsar; 4 binaries at $z=0.02$, with individual SNR in the range $\sim 15-55$;
\item {\bf Dataset3}: 50 MSP; rms noise $50$ns in each pulsar; 8 binaries at $z=0.03$, with individual SNR in the range $\sim 10-40$;
\item {\bf Dataset4}: 50 MSP; rms noise of each pulsar randomly drawn in the range $30-200$ns; 3 binaries at $z=0.01$, with individual SNR in the range $\sim 30-40$.
\end{itemize}
Datasets are in order of increasing complexity (more sources, lower SNR). In the last dataset we tested the algorithm performance when combining time series with 
different noise levels. Sample time series extracted from Dataset3 and Dataset4 are visualized in figure \ref{timeseries}, where we can appreciate the variety of 
imprints depending on the pulsar location in the sky relative to each individual source. 

\section{Results and discussion}

\begin{figure*}
\centering
\begin{tabular}{cc}
\includegraphics[width=0.44\textwidth, keepaspectratio=true, clip=true, viewport=130 450 480 700]{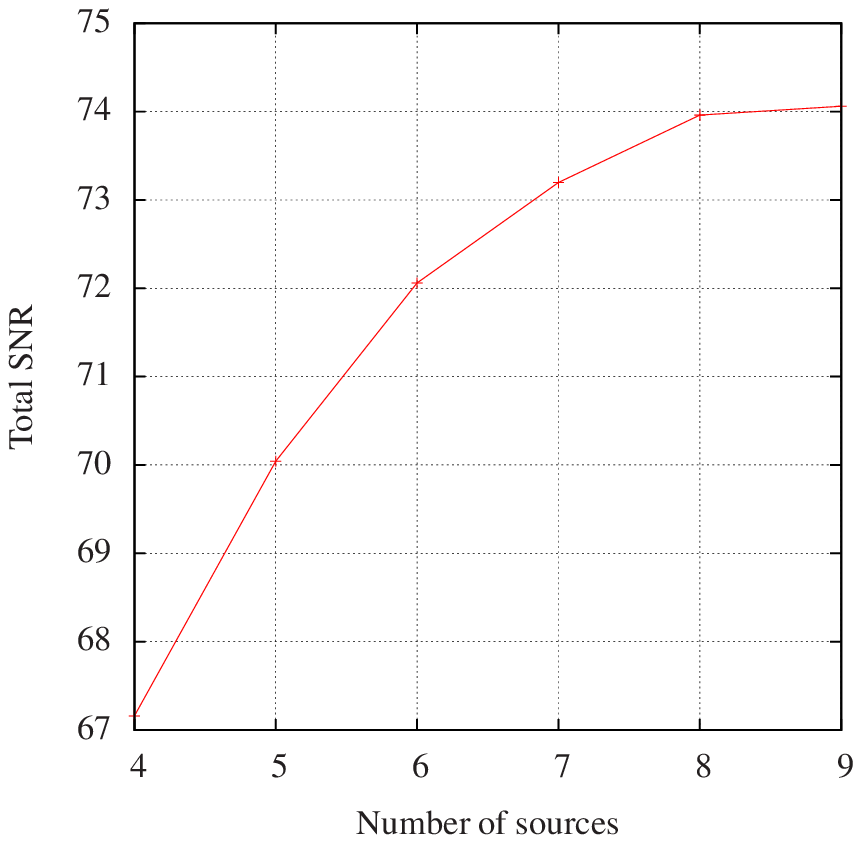} &
\includegraphics[width=0.44\textwidth, keepaspectratio=true, clip=true, viewport=130 450 480 700]{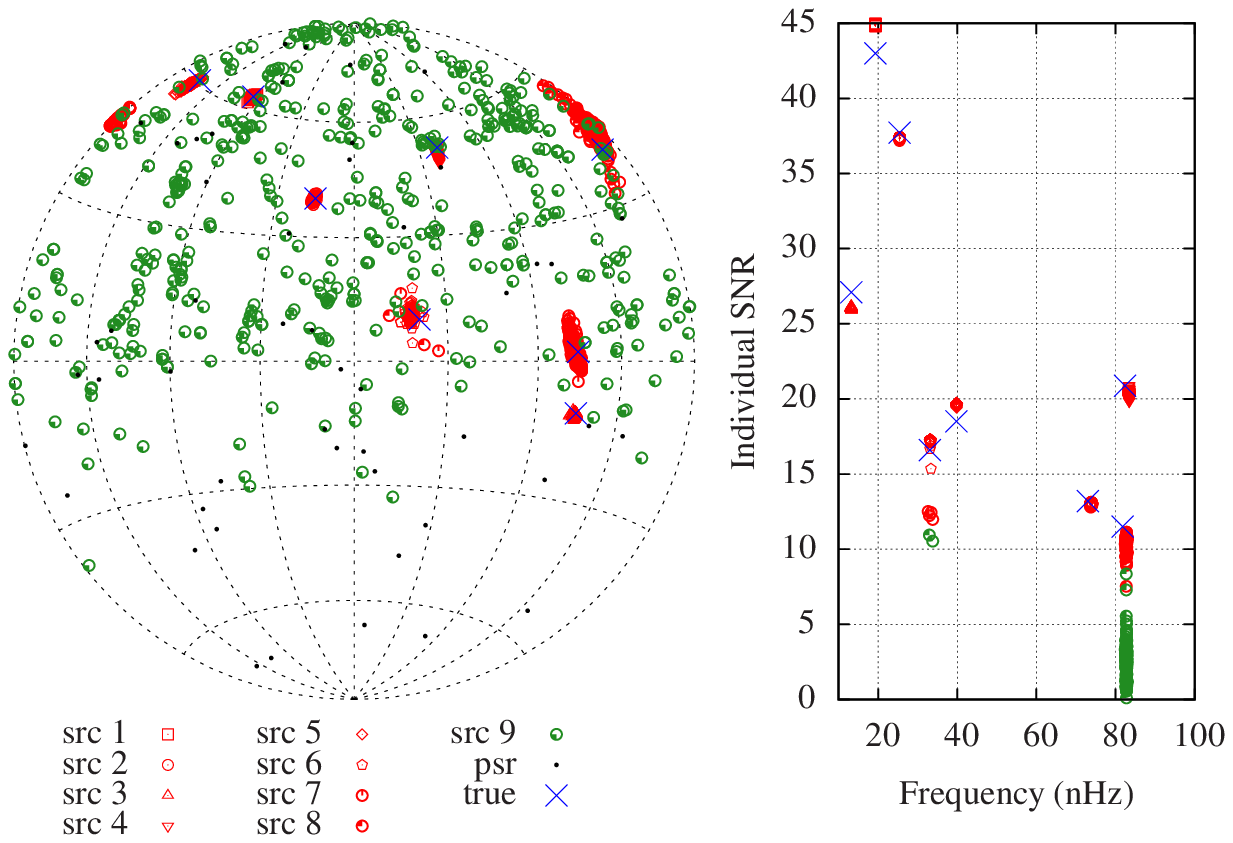} 
%
\end{tabular}
\caption{Performance of the MS-GA in finding the number of sources. Left panel: signal SNR as a function of the number of sources $N_S$ assumed in the template. Right panel: source localization in the sky (sky map) and in the frequency-SNR space (rightmost panel). Blue crosses ($\times$) correspond to injected values and black dots to the position of the MSPs forming the array. Red marks represent sources 1-to-8 found by all the organisms with $\text{SNR}^2_\text{tot} > 99 \% \ \text{SNR}^2_\text{best}$, while green circles represent the 9th source. Note that this latter one does not have a defined position, and typically has SNR$<5$.
}
\label{ResData3src9}
\end{figure*} 

The datasets were generated separately by A. Sesana and were blindly analyzed by A. Petiteau and S. Babak.
The MS-GA was applied to all datasets, adding sources one by one to the template. By doing 
this, we could test the effectiveness of the code in determining both the number of sources in the dataset
and their sky location. A summary of the results is given in table \ref{T:numbers}.

\begin{table*}
\centering
\caption{\label{T:numbers} Recovered and (injected) parameter values of all the simulated sources in each dataset. The last column represent the sky offset of the recovered sources with respect to the injection (see text for details).}
\begin{tabular}{@{}cccccc@{}}
\hline
\hline
&$\,\,\,\,\,\,\,\,\,\,\,\,\,\,\,\,$SNR$\,\,\,\,\,\,\,\,\,\,\,\,\,\,\,\,$ & $\,\,\,\,\,\,\,\,\,\,\,\,\,\,\,\,$$f$[ns]$\,\,\,\,\,\,\,\,\,\,\,\,\,\,\,\,$ & $\,\,\,\,\,\,\,\,\,\,\,\,\,\,\,\,$$\theta$[rad]$\,\,\,\,\,\,\,\,\,\,\,\,\,\,\,\,$ & $\,\,\,\,\,\,\,\,\,\,\,\,\,\,\,\,$$\phi$[rad]$\,\,\,\,\,\,\,\,\,\,\,\,\,\,\,\,$ & $\,\,\,\,\,\,\,\,\Delta\Theta$[deg]$\,\,\,\,\,\,\,\,$\\
\hline
\multirow{5}{*}{{\bf Dataset1}} 
& 60.70 (61.11) & 56.6 (56.4) & 1.249 (1.237) & 2.604 (2.601) & 0.706\\
& 45.85 (42.48) & 38.0 (38.0) & 1.750 (1.748) & 3.765 (3.764) & 0.127\\
& 43.71 (40.43) & 36.4 (36.4) & 1.555 (1.529) & 1.722 (1.712) & 1.596\\
& 35.67 (36.14) & 53.7 (53.5) & 0.537 (0.534) & 5.522 (5.451) & 2.085\\
& 32.27 (31.33) & 48.3 (48.0) & 1.286 (1.295) & 5.144 (5.123) & 1.266\\
\hline
\multirow{4}{*}{{\bf Dataset2}} 
& 54.64 (54.07) & 18.88 (18.9) & 1.774 (1.774) & 3.839 (3.841) & 0.112\\
& 48.01 (47.24) & 11.25 (11.3) & 1.870 (1.858) & 5.720 (5.718) & 0.696\\
& 13.64 (13.05) & 77.42 (76.5) & 0.617 (0.651) & 6.158 (6.115) & 2.434\\
& 12.23 (12.78) & 57.19 (57.0) & 1.613 (1.549) & 6.050 (6.048) & 3.669\\
\hline
\multirow{8}{*}{{\bf Dataset3}} 
& 44.91 (42.99) & 19.33 (19.3) & 0.474 (0.468) & 1.450 (1.454) & 0.359\\
& 37.39 (37.72) & 25.42 (25.4) & 0.883 (0.878) & 2.733 (2.749) & 0.763\\
& 26.02 (27.09) & 13.21 (13.2) & 1.769 (1.764) & 5.078 (5.087) & 0.581\\
& 20.19 (20.88) & 83.42 (82.4) & 0.689 (0.668) & 4.133 (4.162) & 1.593\\
& 19.67 (18.51) & 39.79 (39.8) & 0.541 (0.509) & 0.386 (0.429) & 2.211\\
& 17.27 (16.59) & 33.16 (33.1) & 1.381 (1.397) & 3.621 (3.693) & 4.160\\
& 13.07 (13.19) & 73.83 (73.0) & 1.534 (1.536) & 5.054 (5.078) & 1.379\\
& 10.66 (11.51) & 82.75 (81.8) & 0.809 (0.864) & 6.182 (6.085) & 5.192\\
\hline
\multirow{3}{*}{{\bf Dataset4}} 
& 42.73 (43.92) & 98.2 (96.3) & 2.028 (2.043) & 0.977 (0.961) & 1.200\\
& 28.62 (29.28) & 91.5 (90.1) & 2.655 (2.661) & 1.174 (1.121) & 1.454\\
& 27.56 (28.28) & 48.2 (48.1) & 1.231 (1.245) & 5.774 (5.769) & 0.827\\
\hline
\end{tabular}
\end{table*}

For each dataset we evolved several colonies of organisms assuming $N_s=1, 2, 3,....$ in the template,
we computed the SNR of the best organism at the end of each search, and track its evolution with $N_s$. 
Results are shown in the left panel of figure \ref{ResData3src9} for Dataset3. The maximum SNR steadily
increases by adding sources up to $N_s=8$. Adding a ninth source to the template does not significantly improve 
the match with the data, indicating that the dataset is best described by an eight-source model; in fact 
there were eight sources in Dataset3. The algorithm identified the correct number of sources in all datasets. 
We stress here that all the injected sources had SNR$>10$, high enough to be dug out of the noise. 
In presence of many low SNR sources, we do not expect any search algorithm to recover the correct
number of binaries, but only to identify the brightest ones. We will address this 'confusion problem' in 
a future paper. A complementary view of this result is given in the right panel of figure \ref{ResData3src9}.
We ran several GAs using nine-source colonies of organisms, we identified the solution
(organism) with highest SNR (SNR$_{\rm best}$) and we stored all the organism having $\text{SNR}^2_\text{tot} > 99 \% \ \text{SNR}^2_\text{best}$. 
The figure show the location in the sky
of all sources found in all these 'best solutions'. The location of sources 1-to-8 does not change much
for different solutions, and it is generally consistent with the true (blue crosses) locations of the injected
sources. Conversely, the 9th source (green circles) is extremely scattered around the sky. Moreover, the frequency-SNR plot at the extreme right shows that the individual SNR of
those 9th sources are almost always $<5$, compatible with noise fluctuations.  
  
\begin{figure}
\centering
\begin{tabular}{c}
\includegraphics[width=0.44\textwidth, keepaspectratio=true, clip=true, viewport=130 450 480 700]{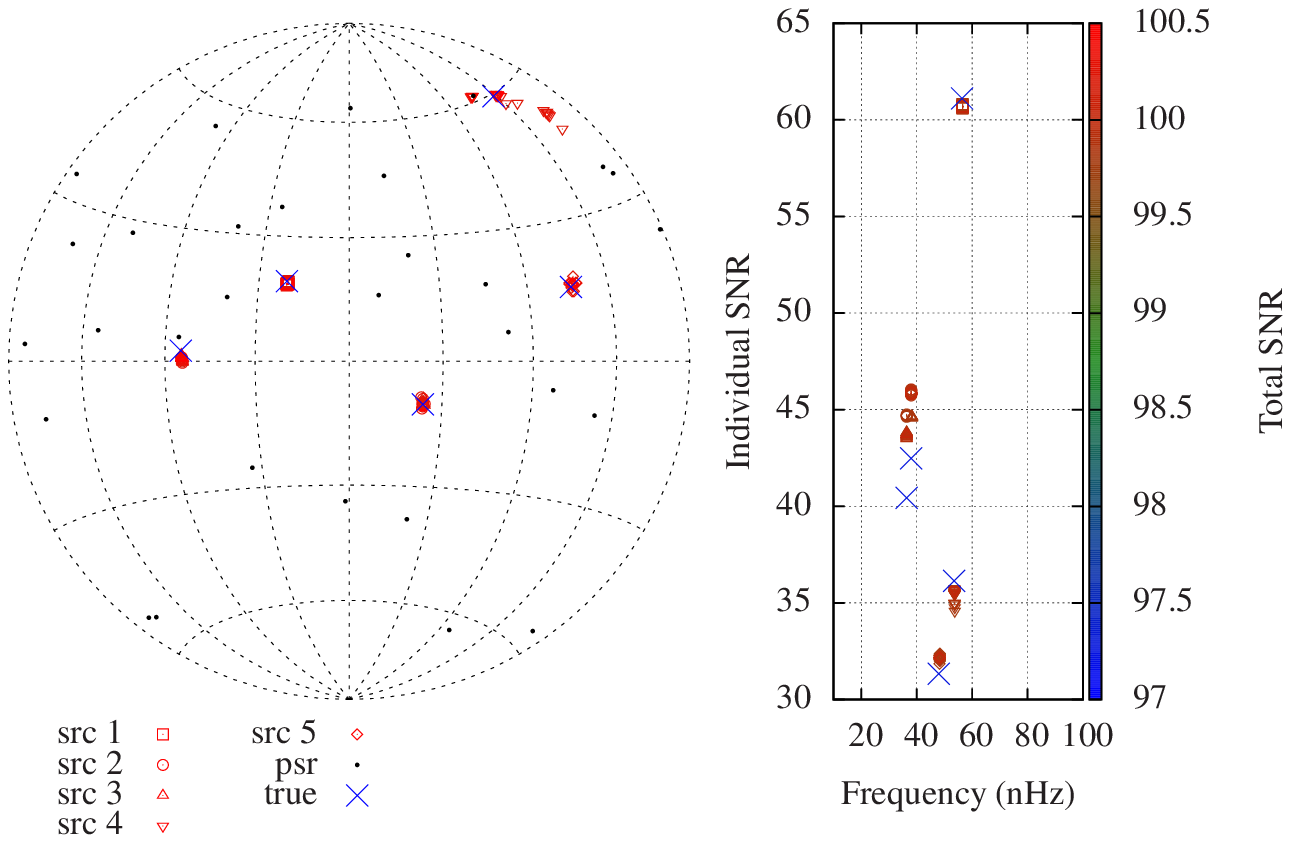} \\
\includegraphics[width=0.44\textwidth, keepaspectratio=true, clip=true, viewport=130 450 480 700]{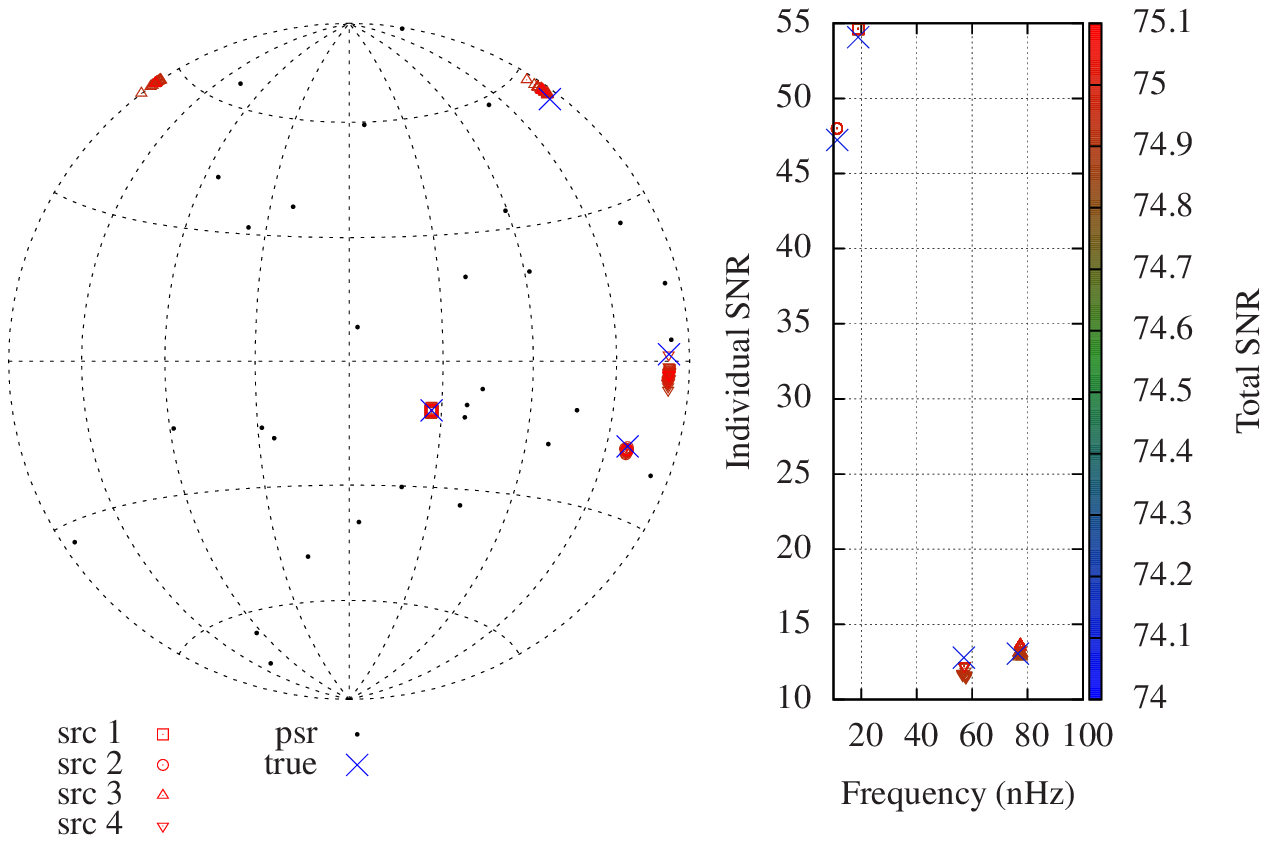}\\
\includegraphics[width=0.44\textwidth, keepaspectratio=true, clip=true, viewport=130 450 480 700]{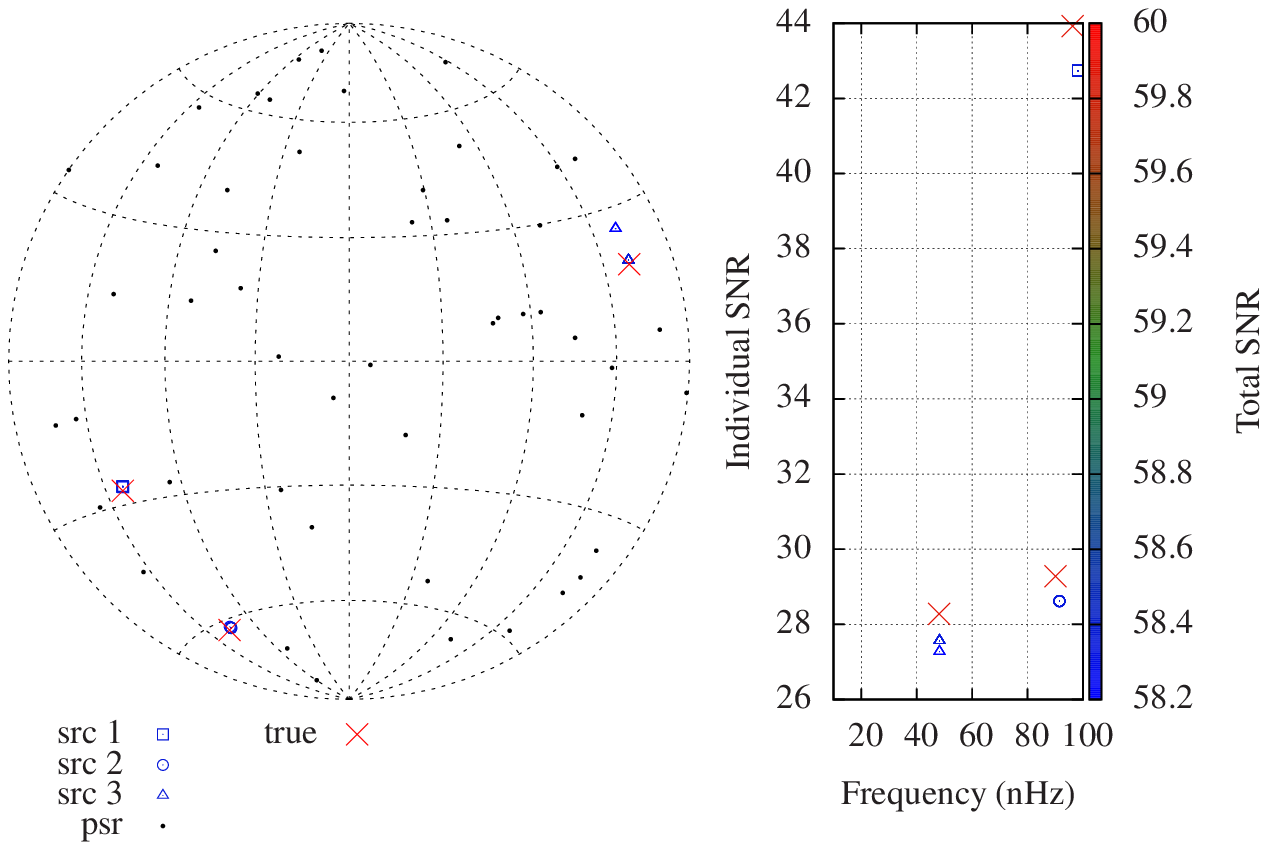}
%
\end{tabular}
\caption{Best solutions for Dataset1 (top panel), Dataset2 (central panel) and Dataset4 (bottom panel). In each panel all solutions with $\text{SNR}^2_\text{tot} > 99.5 \% \ \text{SNR}^2_\text{best}$ are shown. Symbols have the same meaning as in the right panel of Figure~\ref{ResData3src9}. All recovered sources are color-coded according to the rightmost scale, based to the total SNR of the solution they belong to.
}
\label{ResData124SelModes}
\end{figure} 

\begin{figure}
\centering
\begin{tabular}{c}
\includegraphics[width=0.44\textwidth, keepaspectratio=true, clip=true, viewport=130 450 480 700]{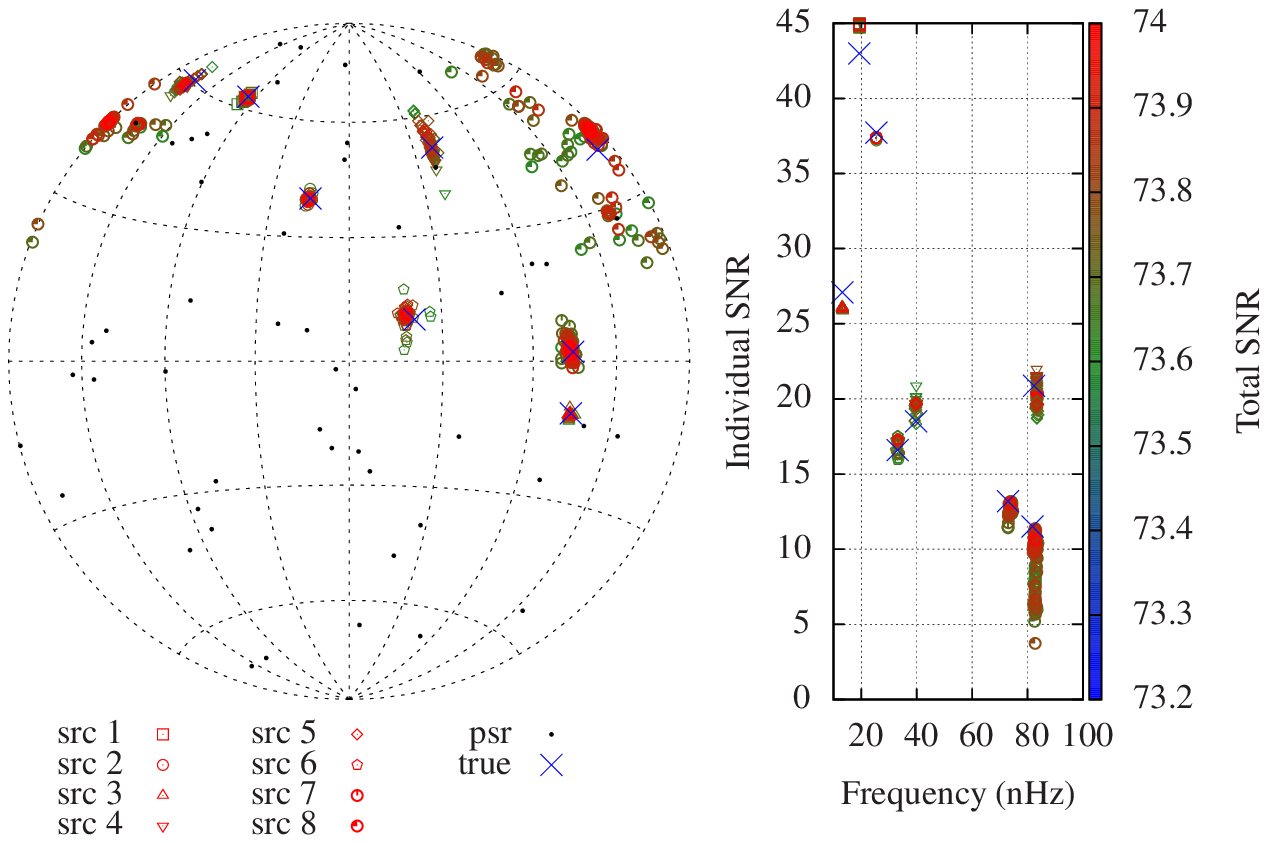} \\
\includegraphics[width=0.44\textwidth,keepaspectratio=true, clip=true, viewport=130 450 480 700]{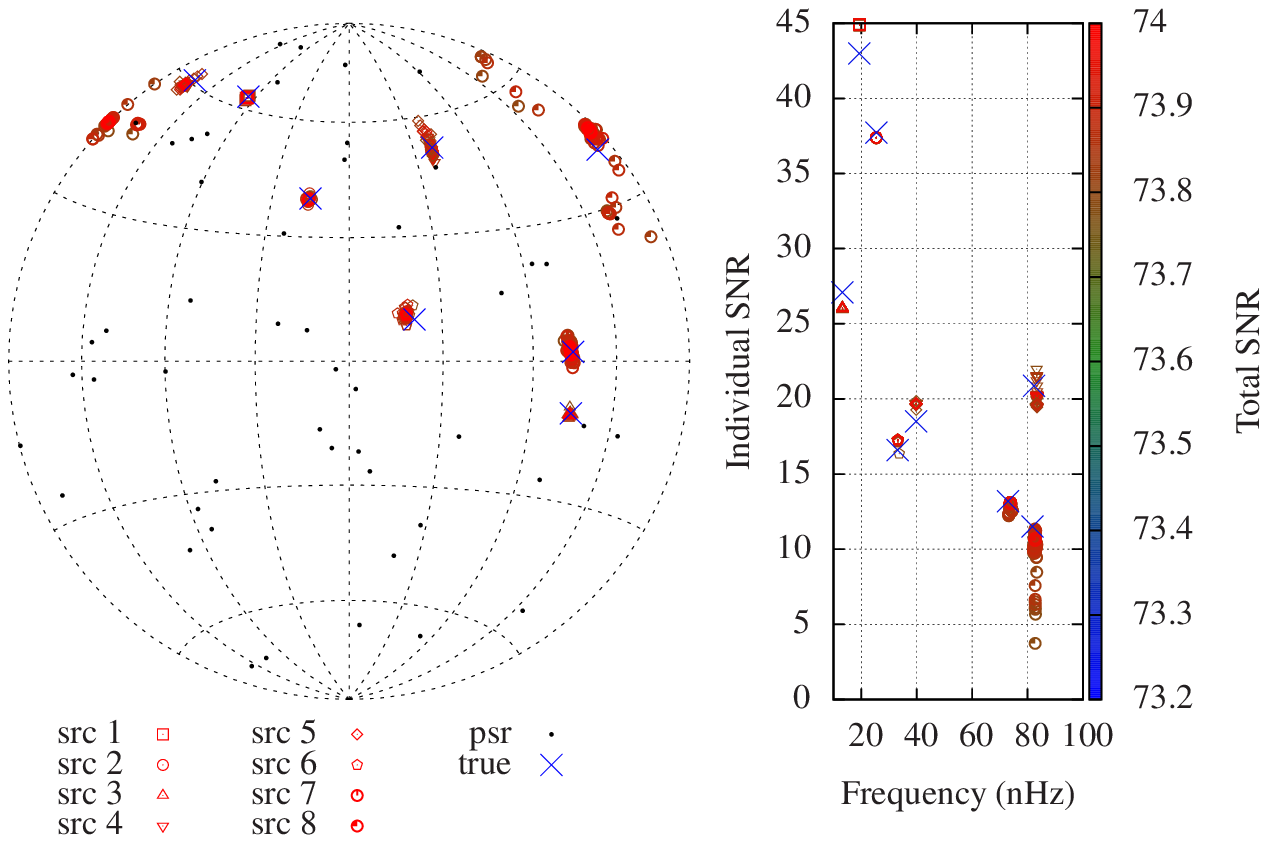}\\
\includegraphics[width=0.44\textwidth, keepaspectratio=true, clip=true, viewport=130 450 480 700]{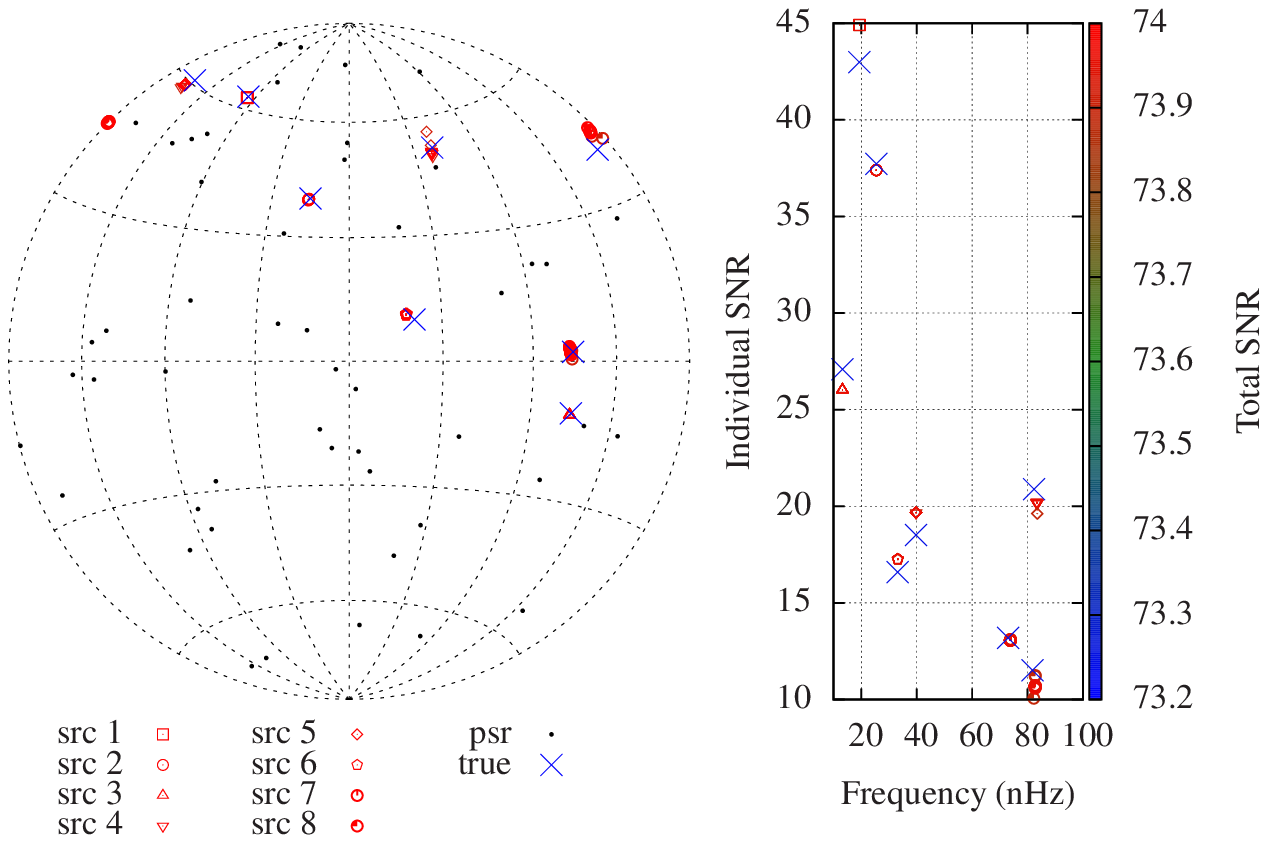}
%
\end{tabular}
\caption{Best solutions for Dataset3. 
The top panel shows all solutions with $\text{SNR}^2_\text{tot} > 99 \% \ \text{SNR}^2_\text{best}$, 
the central panel all solutions with $\text{SNR}^2_\text{tot} > 99.5 \% \ \text{SNR}^2_\text{best}$ 
and the bottom panel all solutions with $\text{SNR}^2_\text{tot} > 99.8 \% \ \text{SNR}^2_\text{best}$. Symbols have the same meaning as in the right panel of Figure~\ref{ResData3src9}. All recovered sources are color-coded according to the rightmost scale, based to the total SNR of the solution they belong to.
}
\label{ResData3SelModes}
\end{figure}

Having tested the code effectiveness in finding the number of sources present in the data, we turn 
now to the description of the results obtained on the individual datasets. Best solutions (those 
with $\text{SNR}^2_\text{tot} > 99.5 \% \ \text{SNR}^2_\text{best}$) for Dataset1 are shown in the 
top panel of figure \ref{ResData124SelModes}. All the five injected sources were found at 
approximately the right sky location, with the right frequency and SNR. Our GA is designed to 
find the modes corresponding to maxima in the likelihood function, but not to explore the exact shape of the likelihood function around those modes. The lack of parameter space exploration around the maxima prevents us to attach fully meaningful errors to our best solutions. We plan to include systematic exploration of the maxima in future work, here we just estimate the sky location error as the angular offset between the best solution and the injected signal. This is defined as $\Delta{\Theta}={\rm arccos}(\vec{n}_t\cdot\vec{n}_r)$, where $\vec{n}_t$ and $\vec{n}_r$ are the unit vectors defining the true sky location of the sources and the recovered value respectively. This is reported in the last column of table \ref{T:numbers}. All sources in Dataset1 are offset by less than 2deg. Results for Dataset2 are shown in the central panel of figure \ref{ResData124SelModes}. 
Again, we see that all sources are correctly identified, despite two of them having SNR just above 10. Sky location offsets $\Delta{\Theta}$ are less than 1deg for the two low frequency bright sources, but degrade to $\sim$3deg for the high frequency, faint ones. Dataset3 was the richest of all, with eight injected sources. Best solutions are shown in figure \ref{ResData3SelModes}, for different SNR threshold, to give a sense of 'how fast' points cluster toward the maximum of the likelihood. Also in this case, all sources are well located in the sky, with brighter sources located better. Looking at table \ref{T:numbers}, we notice that we tend to overstimate the frequencies of sources above $60$ nHz. This is because at such high $f$, the $10^9\msun$ chirp mass binaries injected in the data chirp significantly over the 10 year duration of the observations. Since we are matching the signal with non-evolving monochromatic templates, the estimated frequency is higher than the one injected at the beginning of the observation. However, such high frequency sources seem to have sky location offsets similar (maybe a little worse) to their low frequency counterparts with comparable SNR; i.e., the frequency mismatch between the signal and the template does not seem to corrupt the sky location performance of the search. The same effect is seen in Dataset4 (bottom panel of figure 
\ref{ResData124SelModes} and table \ref{T:numbers}). Also in this case we find source offsets within $\sim$1deg of their true position, but we give a couple of extra nHz to the high frequency sources. Different noise levels in the pulsars do not affect the performance of our search algorithm.

Overall, our MS-GA performed well on all datasets, recovering all the injected sources without returning any false positive. The parameters of the recovered sources well matched the injections
with: (i) sky location offsets less than few degrees, (ii) individual source SNR estimations within few \% 
of the true ones, and (iii) sub-Fourier bin frequency accuracy (sometimes within 0.1 nHz for low frequency sources).
Without a complete exploration of the likelihood function around the maxima, it is difficult to asses proper errors 
on the parameters. We can however take sky position offsets as a proxy of the sky localization accuracy. In fact,
offsets shown in the last column of table \ref{T:numbers} scale (with a large scatter) with the inverse of the SNR. 
This has to be expected: an offset scaling with 1/SNR implies an area of uncertainty scaling with 1/SNR$^2$, in agreement
with theoretical expectations. If we approximate the errorbox in the sky as $\Delta{\Omega}\approx\pi[\Delta{\Theta}]^2$,
we get values in the range 10-70 deg$^2$ for sources with SNR in the range 11-13. This is broadly consistent with
\citep{SesanaVecchio10} who estimated an average sky location accuracy of $\Delta{\Omega}\approx50$deg$^2$ for a source 
observed by an array of 50 pulsars, randomly located in the sky, with total SNR$=10$ (in the Earth term).
Another interesting fact is the frequency mismatch for sources approaching 10$^{-7}$ Hz, caused by their frequency
evolution over the observing time (10 years). This means that, in principle, for such sources, we can measure 
the frequency drift $\dot{f}$, i.e., the chirp rate. In fact, it takes only an extra parameter in the template,
with some extra computational cost. The measure of $\dot{f}$ breaks the chirp mass/luminosity distance degeneracy 
in the source amplitude, allowing for a direct measurement of the source luminosity distance. This, ultimately, 
will narrow down significantly the number of candidate electromagnetic counterparts in the source sky errorbox, 
facilitating a positive identification. However, evolution on such short timescales is detectable only for
very massive (${\cal M}\sim10^9\msun$, as the systems injected in the data) binaries emitting at frequencies
higher than $\sim 7\times10^{-8}$Hz. Intrigued by this possibility, we checked how likely is to find such
extreme systems in realistic populations of MBH binaries in the Universe. We took the models investigated by
\cite{sesana09} and computed the average number of expected sources with ${\cal M}>10^9\msun$ 
and $f>7\times10^{-8}$. Depending on the adopted MBH mass-bulge relation and on the accretion implementation
(see \cite{sesana09} for details), we found average number of sources ranging from $10^{-3}$ to 0.04, i.e., there
is less than 5\% chance to have a such bright high frequency source in the sky. If we relax the mass requirement
to ${\cal M}>5\times10^8\msun$, figures grow to 0.01-0.4. To properly quantify the probability of measuring
$\dot{f}$, one should estimate its minimum measurable value for a given array, and then select in the MBH binary
population all the sources occupying the portion of the chirp mass-frequency parameter space compatible with such value. 
The crude figures estimated here indicate that $\dot{f}$ measurements using the Earth term only should be unlikely.

\section{Conclusions}
This is a second in a series of paper devoted to the exploration of the PTA potential of resolving
multiple GW sources. In Paper~I we addressed basic issues like the number of sources per frequency
bin that can be resolved by an array of $N$ pulsars, demonstrating our findings with primitive searches
on several (mostly noiseless) synthetic datasets. Here we pushed our analysis a bit further by (i)
extending the mathematical formulation of the likelihood function to include the source frequencies
as additional free parameters and by (ii) implementing a multi-search genetic algorithm to efficiently find
the maximum of the likelihood function.

We constructed synthetic datasets consisting of collections of time series representing the 
residuals obtained by timing an ensemble of MSPs. MSPs were placed randomly in the sky, 
each time series consisted of 523 equally sampled datapoints over an observing time of 10 
years (one datapoint every two weeks), and the noise in the data was assumed to be white Gaussian. In each
dataset we injected an unknown number $N_S$ of sources with random parameters and individual
$\text{SNR}>10$ and we apply our multi-search genetic algorithm to search for their sky location and
frequency. Note that we assumed circular monochromatic sources in our template, but we allowed for
full PN evolution of the injected sources. By doing so, we placed ourselves in the (likely) situation
in which the theoretical model of the signal does not perfectly represent its real nature, and we
explored the consequences of this mismatch.

Our main results can be summarized as follows:
\begin{itemize}
\item the MS-GA generally converged to the true maximum of the likelihood function in 2-to-5
iterations (few hours on one core at  2GHz);
\item the MS-GA successfully identified all the injected sources in all datasets. No false positive
were found;
\item the search on all source parameters was successful: inferred sky locations were offset by less than few degrees,
individual source SNR estimations matched the injections within few \%, and frequencies were determined with sub-Fourier bin precision (most of the times to better than 0.1nHz);
\item sky location offsets roughly scaled with 1/SNR implying a sky location accuracy scaling as 1/SNR$^2$. Even though we did not compute proper errorboxes in the sky, we estimated source localization capabilities broadly consistent with theoretical expectations derived in \cite{SesanaVecchio10} under similar assumptions;
\item we overestimated the frequency of sources approaching $f=10^{-7}$Hz. This is because massive systems at such high frequency significantly chirp during the observation time (whereas chirp was not allowed in our template). This means that we can measure $\dot{f}$ and therefore estimate the chirp mass and, in turn, the luminosity distance of the source. Although this is a very appealing prospect, we estimated on average less than 1 source with a measurable $\dot{f}$ in a realistic realization of the MBH binary population in the Universe;
\item the MS-GA performances do not seem to be affected by unequal noise levels in different MSPs.
\end{itemize}

Our results are encouraging, however, they were still obtained under a number of simplifying assumptions that we wish to relax in our future work. Firstly, datasets were still evenly sampled, with no gaps; an idealised situation that is not going to occur in reality. Secondly, we just took noisy datastreams and fit for the GW sources only, implicitly assuming perfectly known MSP parameters; any realistic detection pipeline must fit for MSP parameters and GW signals simultaneously. Finally, we still did not include the pulsar terms in our injections; those are likely to blend together with lower frequency Earth terms to bias estimated source parameters and to (maybe) create false positives. Only by relaxing those assumptions we will be able to demonstrate the effectiveness of our MS-GA algorithm in tackling a problem with realistic complexity. We plan to investigate these issues in the next paper of the series. We will then try to apply our search algorithm to raw times of arrival, carrying the imprint of a realistic population of MBH binaries.

\section*{Acknowledgments}

Work of A.S. and S.B. was supported in parts by DFG grant SFB/TR 7 Gravitational Wave Astronomy and by DLR
(Deutsches Zentrum fur Luft- und Raumfahrt).

\bibliographystyle{h-physrev4}
\bibliography{references}

\end{document}